\documentstyle[12pt,amsmath,myart,graphicx]{article}

\begin{document}
\bigskip

\hfill\hbox{}

\bigskip

\begin{center}
{\Large \textbf{Boundary Liouville Field Theory}}

{\Large \textbf{I. Boundary State and Boundary Two-point Function}}

\vspace{0.8cm}

{\large V.Fateev}

\vspace{0.5cm} {\large A.Zamolodchikov}\footnote{Department of Physics and
Astronomy, Rutgers University, P.O.Box 849, Piscataway, New Jersey 08855-0849,
USA and L.D.Landau Institute for Theoretical Physics, ul.Kosygina 2, 117334, 
Moscow, Russia}

\vspace{0.3cm}

and

\vspace{0.5cm} {\large Al.Zamolodchikov}

\vspace{0.2cm}

Laboratoire de Physique Math\'ematique\footnote{\textrm{Laboratoire Associ\'e
au CNRS URA 768}}

Universit\'e Montpellier II

Place E.Bataillon, 34095 Montpellier, France
\end{center}

\textbf{Abstract}

Liouville conformal field theory is considered with conformal boundary. There
is a family of conformal boundary conditions parameterized by the boundary
cosmological constant, so that observables depend on the dimensional ratios of
boundary and bulk cosmological constants. The disk geometry is considered. We
present an explicit expression for the expectation value of a bulk operator
inside the disk and for the two-point function of boundary operators. We comment
also on the properties of the degenrate boundary operators. Possible
applications and further developments are discussed. In particular, we present exact
expectation values of the boundary operators in the boundary sin-Gordon model.

\section{Liouville field theory}

During last 20 years the Liouvlle field theory permanently attracts much
attention mainly due to its relevance in the quantization of strings in
non-critical space-time dimensions \cite{Polyakov} (see also refs.\cite{CT,
GN, HJ}). It is also applied as a field theory of the 2D quantum gravity.
E.g., the results of the Liouville field theory (LFT) approach can be compared
with the calculations in the matrix models of two-dimensional gravity
\cite{Kazakov1, Brezin} and this comparison shows \cite{KPZ, David} that when
the LFT central charge $c_{L}\geq25$ this field theory describes the same
continuous gravity as was found in the critical region of the matrix models.
Although there are still no known applications of LFT with $c_{L}<25$, the
theory is interesting on its own footing as an example of non-rational 2D
conformal field theory.

In the bulk the Liouville field theory is defined by the Lagrangian density
\begin{equation}
{\mathcal{L}}
=\frac{1}{4\pi}(\partial_{a}\phi)^{2}+\mu e^{2b\phi} \label{Liouv}%
\end{equation}
where $\phi$ is the two-dimenesional scalar field, $b$ is the dimensionless
Liouville coupling constant and the scale parameter $\mu$ is called the
cosmological constant. This expression implies a trivial background metric
$g_{ab}=\delta_{ab}$. In more general background the action reads
\begin{equation}
A_{\mathrm{bulk}}=\frac{1}{4\pi}\int\limits_{\Gamma}\left[  g^{ab}\partial
_{a}\phi\partial_{b}\phi+QR\phi+4\pi\mu e^{2b\phi}\right]  \sqrt{g}d^{2}x
\label{liouv}%
\end{equation}
Here $R$ is the scalar curvature associated with the background metric $g$
while $Q$ is an important quantity in the Liouville field theory called the
background charge
\begin{equation}
Q=b+1/b \label{Q}%
\end{equation}
It determines in particular the central charge of the theory
\begin{equation}
c_{L}=1+6Q^{2} \label{charge}%
\end{equation}

In what follows we always will consider only the simplest topologies like
sphere or disk which can be described by a trivial background. For example, a
sphere can be represented as a flat projective plane where the flat Liouville
lagrangian (\ref{Liouv}) is valid if we put away all the curvature to the
spacial infinity where it is seen as a special boundary condition on the
Liouville field $\phi$
\begin{equation}
\phi(z,\bar{z})=-Q\log(z\bar{z})+O(1)\ \ \ \ \ \mathrm{at}\ \ |z|\rightarrow
\infty\label{sphere}%
\end{equation}
called the background charge at infinity.

The basic objects of LFT are the exponential fields $V_{\alpha}(x)=\exp
(2\alpha\phi(x))$ which are conformal primaries w.r.t. the stress tensor
\begin{align}
T(z{)}  &  {}=-(\partial\phi)^{2}+Q\partial^{2}\phi\label{stress}\\
\bar{T}(\bar{z}{)}  &  {}=-(\bar{\partial}\phi)^{2}+Q\bar{\partial}^{2}%
\phi\nonumber
\end{align}
The field $V_{\alpha}$ has the dimension
\begin{equation}
\Delta_{\alpha}=\alpha(Q-\alpha) \label{dim}%
\end{equation}
In fact not all of these operators are independent. One has to identify the
operators $V_{\alpha}$ and $V_{Q-\alpha}$ so that the whole set of local LFT
fields is obtained by the ``folding'' of the complex $\alpha$-plane w.r.t.
this reflection. The only exception is the line $\alpha=Q/2+iP$ with $P$ real
where these exponential fields, if interpreted in terms of quantum gravity,
seem not to correspond to local operators. E.g. in the classical theory they
appear as hyperbolic solutions to the Lioville equation and ``create holes''
in the surface \cite{seiberg, polch}. Instead, these values of $\alpha$ are
attributed to the normalizable states. The LFT space of states $\mathcal{A}$
consists of all conformal families $[v_{P}]$ corresponding to the primary
states $\left|  v_{P}\right\rangle $ with real $0\leq P<\infty$, i.e.,
\begin{equation}
{\mathcal{A}}
=\bigotimes_{P\in\lbrack0,\infty)}[v_{P}] \label{physspace}%
\end{equation}
The primary states $\left|  v_{P}\right\rangle $ are related to the values
$\alpha=Q/2+iP$ and have dimensions $Q^{2}/4+P^{2}$ while other values of
$\alpha$ are mapped onto non-normalizable states. This is a peculiarity of the
operator-state correspondence of the Liouville field theory which differs it
from conventional CFT with discret spectra of dimensions but make it similar
to some conformal $\sigma$-models with non-compact target spaces. In what
follows the primary physical states are normalized as
\begin{equation}
\left\langle v_{P^{\prime}}|v_{P}\right\rangle =\pi\delta(P-P^{\prime})
\label{norm}%
\end{equation}

The solution of the spherical LFT amounts to constructing all multipoint
correlation functions of these fields,
\begin{equation}
G_{\alpha_{1\ldots}\alpha_{N}}(x_{1},\ldots,x_{N})=\left\langle V_{\alpha_{1}%
}(x_{1})\ldots V_{\alpha_{N}}(x_{N})\right\rangle \label{multipoint}%
\end{equation}
In principle these quantities are completely determined by the structure of
the operator product expansion (OPE) algebra of the exponential operators,
i.e. can be completely restored from the two-point function%

\begin{equation}
\left\langle V_{a}(x)V_{a}(0)\right\rangle =\frac{D(\alpha)}{(x\bar
{x})^{2\Delta_{\alpha}}} \label{twopoint}%
\end{equation}
which determines the normalization of the basic operators and the three-point
function
\begin{equation}
G_{\alpha_{1},\alpha_{2},\alpha_{3}}(x_{1},x_{2},x_{3})=\left|  x_{12}\right|
^{2\gamma_{3}}\left|  x_{23}\right|  ^{2\gamma_{1}}\left|  x_{31}\right|
^{2\gamma_{2}}C(\alpha_{1},\alpha_{2},\alpha_{3}) \label{point3}%
\end{equation}
with $\gamma_{1}=\Delta_{\alpha_{1}}-\Delta_{\alpha_{2}}-\Delta_{\alpha_{3}}$,
$\gamma_{2}=\Delta_{\alpha_{2}}-\Delta_{\alpha_{3}}-\Delta_{\alpha_{1}}$,
$\gamma_{3}=\Delta_{\alpha_{3}}-\Delta_{\alpha_{1}}-\Delta_{\alpha_{2}}$ .
Once these quantities are known, the multipoint functions can be in principle
reconstructed by the purely ``kinematic'' calculations relied on the conformal
symmetry only. Although these calculations present a separate rather
complicated technical problem, conceptually one can say that a CFT (on a
sphere) is constructed if these basic objects are found.

For LFT these quantities were first obtained by Dorn and
Otto\cite{Dorn1,Dorn2} in 1992 (see also \cite{AAl}). We will present here the
derivation of the simplest of them, the two-point function $D(\alpha)$, to
illustrate a different approach to this problem proposed more recently by
J.Techner \cite{Teschner} which seems more effecticient. Close ideas are also
developed in the studies of LFT\ by Gervais and collaborators \cite{Gervais}.
We will use similar approach shortly in the discussion of the boundary
Liouville problem.

Among the exponential operators $V_{\alpha}$ there is a series of fields
$V_{-nb/2}$, $n=0,1,\ldots$ which are degenerate w.r.t. the conformal symmetry
algebra and therefore satisfy certain linear differential equations. For
example, the first non-trivial operator $V_{-b/2}$ satisfies the following
second order equation
\begin{equation}
\left(  \frac{1}{b^{2}}\partial^{2}+T(z)\right)  V_{-b/2}=0 \label{twodiff}%
\end{equation}
and the same with the complex conjugate differentiation in $\bar{z}$ and
$\bar{T}(\bar{z})$ instead of $T$. In the classical limit of LFT the existence
of this degenerate operator can be traced back to the well known relation
between the ordinary second-order linear differential equation and the
classical partial-derivative Liouville equation \cite{Poincare}. The next
operator $V_{-b}$ satisfies two complex conjugate third-order differential
equations
\begin{equation}
\left(  \frac{1}{2b^{2}}\partial^{3}+2T(z)\partial+(1+2b^{2})\partial
T(z)\right)  V_{-b}=0 \label{threediff}%
\end{equation}
and so on. It follows from these equations that the operator product expansion
of these degenerate operators with any primary field, in the present case with
our basic exponential fields $V_{\alpha}$, is of very special form and
contains in the r.h.s only finite number of primary fields. For example for
the first one there are only two representations%

\begin{equation}
V_{-b/2}V_{\alpha}=C_{+}\left[  V_{\alpha-b/2}\right]  +C_{-}\left[
V_{\alpha+b/2}\right]  \label{twoterm}%
\end{equation}
where $C_{\pm}$ are the special structure constants. What is important to
remark about these special structure constants is that the general CFT and
Coulomb gas experience suggests that they can be considered as
``perturbative'', i.e. are obtained as certain Coulomb gas (or ``screening'')
integrals \cite{FF, Dotsenko}. For example in our case in the first term of
(\ref{twoterm}) there is no need of screening insertion and therefore one can
set $C_{+}=1$. The second term requires a first order insertion of the
Liouville interaction $-\mu\int\exp(2b\phi)d^{2}x$ and%

\begin{align}
C_{-}  &  =-\mu\int d^{2}x\left\langle V_{\alpha}(0)V_{-b/2}(1)e^{2b\phi
(x)}V_{Q-\alpha-b/2}(\infty)\right\rangle \nonumber\\
&  =-\mu\dfrac{\pi\gamma(2b\alpha-1-b^{2})}{\gamma(-b^{2})\gamma(2b\alpha)}
\label{Cplus}%
\end{align}
where as usual $\gamma(x)=\Gamma(x)/\Gamma(1-x)$. It is remarkable that all
the special structure constants entering the special truncated OPE's with the
degenerate fields can be obtained in this way.

Now let us take the two-point function $D(\alpha)$ and consider the auxiliary
three-point function%

\[
\left\langle V_{\alpha}(x_{1})V_{\alpha+b/2}(x_{2})V_{-b/2}(z)\right\rangle
\]
Then, tending $z\rightarrow x_{1}$ we see that in the OPE only the second term
survives and in fact our auxiliary function is $\sim C_{-}D(\alpha+b/2)$.
Instead tending $z\rightarrow x_{2}$ we can ``lower'' the parameter of the
second operator down to $\alpha$ which results in $C_{+}D(\alpha)$. Equating
these two things we arrive at the functional equation for the two-point
function
\begin{equation}
\frac{D(\alpha+b/2)}{D(\alpha)}=C_{-}^{-1}(\alpha) \label{DD}%
\end{equation}
This equation can be easily solved in terms of gamma-functions%

\begin{equation}
D(\alpha)=\left(  \pi\mu\gamma(b^{2})\right)  ^{(Q-2\alpha)/b}\frac
{\gamma(2b\alpha-b^{2})}{b^{2}\gamma(2-2\alpha/b+1/b^{2})} \label{Dexplicit}%
\end{equation}
which coincides precisely with what was obtained for this quantity in the
original studies.

In fact there are many solutions to the above functional equation. It is
relevant for the moment to stop at the remarkable duality property of LFT.
Besides the abovementioned series of degenerate operators $V_{-nb/2}$ there is
a ``dual'' series with $b$ replaced by $1/b$. This results in another ``dual''
functional equation for $D(\alpha)$ with the shift by $1/b$ instead of $b$.
The solution becomes unique (at least if these two shifts are uncomparable)
\cite{Teschner}. Note that these two equations are compatible only if in the
dual equation the cosmological constant $\mu$ is replaced by the ``dual
cosmological constant'' $\tilde{\mu}$ related to $\mu$ as follows
\begin{equation}
\pi\tilde{\mu}\gamma(1/b^{2})=\left(  \pi\mu\gamma(b^{2})\right)  ^{1/b^{2}%
}\label{mutwiddle}%
\end{equation}
With this definition of $\tilde{\mu}$ the duality property, which turn out to
hold exactly in LFT, can be formulated as the symmetry of all observables
w.r.t. the substitution $b\rightarrow1/b$ and $\mu\rightarrow\tilde{\mu}$.

The same way one can readily obtain and solve the functional equations for the
three-point function \cite{Teschner} which reads
\begin{align}
\  &  {C}(\alpha_{1},\alpha_{2},\alpha_{3})=\left[  \pi\mu\gamma
(b^{2})b^{2-2b^{2}}\right]  {^{(Q-\sum\alpha_{i})/b}\times}\label{threepoint}%
\\
&  \ \ \frac{\Upsilon_{0}\Upsilon(2\alpha_{1})\Upsilon(2\alpha_{2}%
)\Upsilon(2\alpha_{3})}{\Upsilon(\alpha_{1}+\alpha_{2}+\alpha_{3}%
-Q)\Upsilon(\alpha_{1}+\alpha_{2}-\alpha_{3})\Upsilon(\alpha_{2}+\alpha
_{3}-\alpha_{1})\Upsilon(\alpha_{3}+\alpha_{1}-\alpha_{2})}\nonumber
\end{align}
where a special function $\Upsilon(x)$ has to be introduced
\begin{equation}
\log\Upsilon(x)=\int_{0}^{\infty}\frac{dt}{t}\left[  \left(  Q/2-x\right)
^{2}e^{-2t}-\frac{\sinh^{2}\left(  Q/2-x\right)  t}{\sinh(bt)\sinh
(t/b)}\right]  \label{Upsilon}%
\end{equation}
This integral representation is convergent only in the strip $0<\mathrm{Re}$
$x<Q$, otherwise it is an analytic continuation. In fact $\Upsilon(x)$ is an
entire function of $x$ with zeroes at $x=-nb-m/b$ and $x=Q+nb+m/b$ with $n$
and $m$ non-negative integers.

In the sense mentioned above the explicit results (\ref{twopoint}) and
(\ref{threepoint}) constitute the exact construction of the Liouville field
theory on a sphere. For example, the four-point function can be explicitly
expressed in terms of the three-point function
\begin{equation}
G_{\alpha_{1},\alpha_{2},\alpha_{3},\alpha_{4}}(x_{i})={\frac{1}{\pi}}%
\int\limits_{0}^{\infty}C(\alpha_{1},\alpha_{2},Q/2+iP)C(\alpha_{3},\alpha
_{4},Q/2-iP)|{\mathbf{F}}(\Delta_{\alpha_{i}},\Delta,x_{i})|^{2}dP
\label{fourpoint}%
\end{equation}
where the intergration is over the variety of physical states $\left|
v_{P}\right\rangle $ and ${\mathbf{F}}
(\Delta_{\alpha_{i}},\Delta,x_{i})$ is the
four-point conformal block, determined completely by the conformal symmetry
\cite{BPZ}\footnote{Strictly speaking, (\ref{fourpoint}) is literally correct
only if Re$\alpha_{i}$ are sufficiently close to $Q/2$. Otherwise additional
discrete terms in the r.h.s of (\ref{fourpoint}) can appear due to certain
poles of $C $ braeking through the integration contour, see \cite{BPZ}.}. In
the four-point case, which we are considering now, the latter can be further
reduced to a function of one variable, e.g.,
\begin{align}
{\mathbf{F}}(\Delta_{\alpha_{i}},\Delta,x_{i})  &  =\label{block}\\
&  \ (x_{4}-x_{1})^{-2\Delta_{1}}(x_{4}-x_{2})^{\Delta_{1}+\Delta_{3}%
-\Delta_{2}-\Delta_{4}}(x_{4}-x_{3})^{\Delta_{1}+\Delta_{2}-\Delta_{3}%
-\Delta_{4}}(x_{3}-x_{2})^{\Delta_{4}-\Delta_{1}-\Delta_{2}-\Delta_{3}}%
\times\nonumber\\
&  \ {\mathcal{F}}\left(
\begin{array}
[c]{cc}%
\alpha_{1} & \alpha_{3}\\
\alpha_{2} & \alpha_{4}%
\end{array}
\mid P\mid\eta\right) \nonumber
\end{align}
where
\[
\eta=\frac{(x_{1}-x_{2})(x_{3}-x_{4})}{(x_{1}-x_{4})(x_{3}-x_{2})}
\]
Parameters $\alpha_{i}$ are related to $\Delta_{\alpha_{i}}$ as in
eq.(\ref{dim}) and in the intermediate dimension $\Delta=Q^{2}/4+P^{2}$.

\section{The boundary Liouville problem}

The basic ideas of 2D conformal field theory with conformally invariant
boundary were developed long ago mostly by J.Cardy \cite{Cardy0} who also
applied them successfully to rational CFT's, in particular to the minimal
series \cite{Cardy1, Cardy2}. Here we'll try to apply these ideas to the
Liouville CFT with boundary.

A conformally invariant boundary condition in LFT can be introduced through
the following boundary interaction
\begin{equation}
A_{\mathrm{bound}}=A_{\mathrm{bulk}}+\int\limits_{\partial\Gamma}\left(
\frac{QK}{2\pi}\phi+\mu_{B}e^{b\phi}\right)  g^{1/4}d\xi\label{gbound}%
\end{equation}
where the integration in $\xi$ is along the boundary while $K$ is the
curvature of the boundary in the background geometry $g$. In what follows we
consider only the geometry of a disk which can be represented as a simply
connected domain $\Gamma$ in the complex plane with a flat background metric
$g_{ab}=\delta_{ab}$ inside. The action is simplified as
\begin{equation}
A_{\mathrm{bound}}=\int\limits_{\Gamma}\left(  \frac{1}{4\pi}(\partial_{a}%
\phi)^{2}+\mu e^{2b\phi}\right)  d^{2}x+\int\limits_{\partial\Gamma}\left(
\frac{Qk}{2\pi}\phi+\mu_{B}e^{b\phi}\right)  d\xi\label{bound}%
\end{equation}
where $k$ is the curvature of the boundary in the complex plane. Typically the
most convenient domain is either a unit circle or the upper half-plane. In the
last case the boundary $\partial\Gamma$ is the real axis and one can omit the
term linear in $\phi$ in the boundary action (\ref{bound}). The price is again
a ``background charge at infinity'', i.e., the same boundary condition on the
field $\phi$ at infinity in the upper half plane
\begin{equation}
\phi(z,\bar{z})=-Q\log(z\bar{z})+O(1)\ \ \ \ \ \mathrm{at}\ \ |z|\rightarrow
\infty\label{bcharge}%
\end{equation}
as in the case of the sphere.

It seems natural to call the additional parameter $\mu_{B}$ the boundary
cosmological constant. We see that in fact there is a one-parameter family of
conformally invariant boundary conditions characterized by different values of
the boundary cosmological constant $\mu_{B}$. Contrary to the pure bulk
situation where the cosmological constant enters only as a scale parameter,
the observables in the boundary case actually depend on the scale invariant
ratio $\mu/\mu_{B}^{2}$. For example, a disk correlation function with the
bulk operators $V_{\alpha_{1}},V_{\alpha_{2}}\cdots V_{\alpha_{n}}$ and the
boundary operators (see below) $B_{\beta_{1}},B_{\beta_{2}}\cdots B_{\beta
_{m}}$ scales as follows
\begin{equation}
{\mathcal{G}}(\alpha_{1,}\cdots\alpha_{n},\beta_{1},\cdots\beta_{m})\sim
\mu^{(Q-2\sum_{i}\alpha_{i}-\sum_{j}\beta_{j})/2b}F\left(  \frac{\mu_{B}^{2}%
}{\mu}\right)  \label{scaling}%
\end{equation}
where $F$ is some scaling function and we indicate only the dependence on the
scale parameters $\mu$ and $\mu_{B}$\footnote{In the presence of boundary
operators it is possible to impose different boundary conditions at different
pieces of the boundary, each being characterised by its own value of $\mu_{B}%
$. In this case the scaling function in (\ref{scaling}) may depend on several
invariant ratios, see below.}. Our present purpose is to study this dependence.

In the boundary case we have to introduce the boundary operators. In LFT the
basic boundary primaries are again the exponential in $\phi$ boundary fields
$B_{\beta}=\exp\beta\phi$. Their dimensions are
\begin{equation}
\Delta_{\beta}=\beta\left(  Q-\beta\right)  \label{bdim}%
\end{equation}
To avoid any confusions we shall always use parameter $\alpha$ for the bulk
exponentials and parameter $\beta$ in relation with the boundary operators. In
general a boundary operator is not characterized completely by its dimension,
because the conformal boundary conditions at both sides of the location of the
boundary operator may be in general different. One has to specify which
boundary condition it joins. Therefore in general we are talking about a
juxtaposition boundary operator between, in our case, two boundary conditions
with the parameters $\mu_{B_{1}}=\mu_{1}$ and $\mu_{B_{2}}=\mu_{2}$ and denote
it $B_{\beta}^{\mu_{1}\mu_{2}}(x)$.

To define completely the boundary LFT on the disk, i.e, to be able to
construct an arbitrary multipoint correlation function including bulk and
boundary operators, we have to reveal few more basic objects in addition to
the bulk two- and three-point functions (\ref{Dexplicit}) and
(\ref{threepoint}) we already have.

\begin{figure}
[tbh]
\begin{center}
\includegraphics[
height=4.8847in,
width=5.0099in
]%
{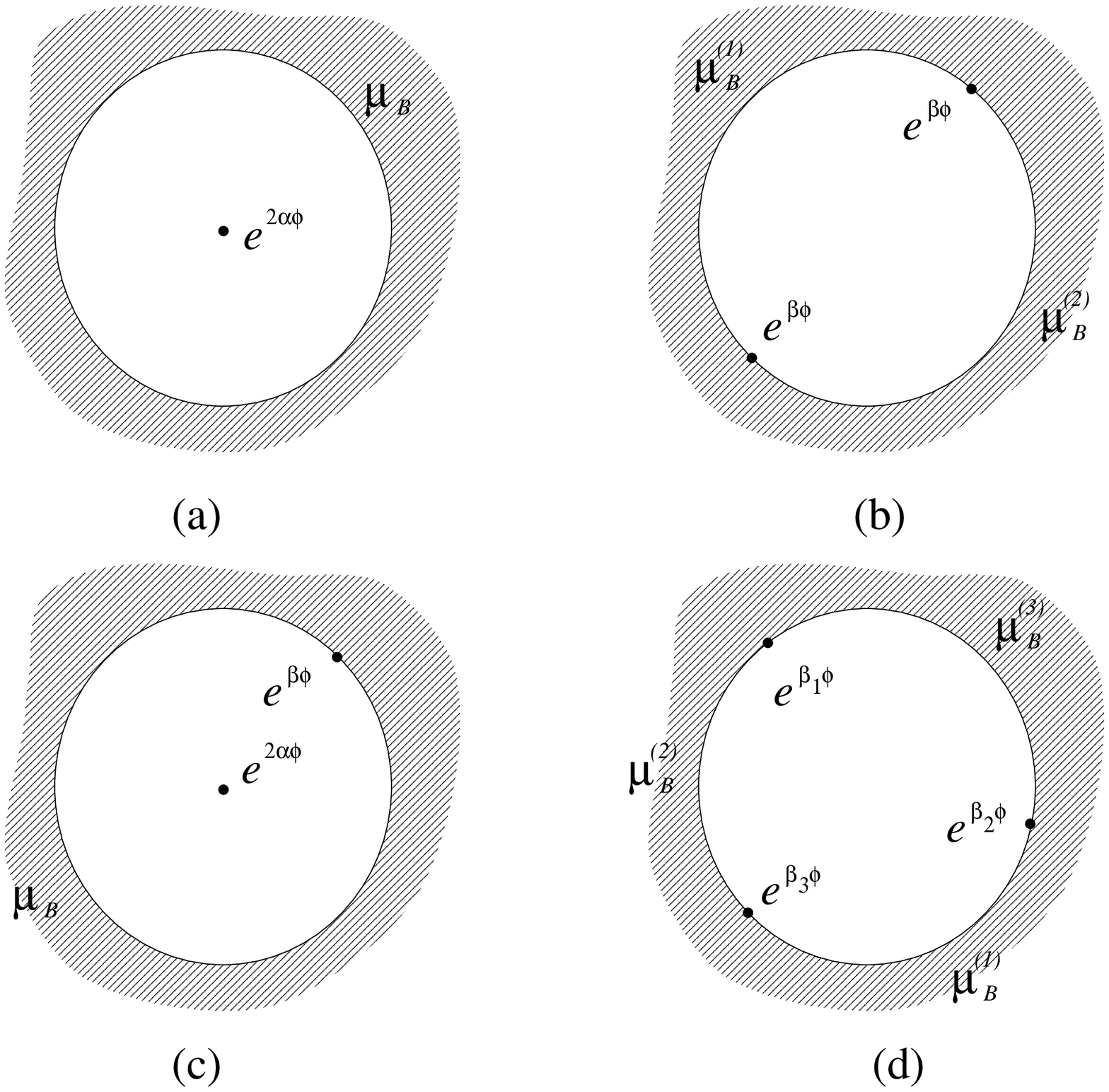}%
\caption{}%
\label{fig1}%
\end{center}
\end{figure}

1. First is the bulk one-point function (we imply almost constantly the upper
half-plane geometry)
\begin{equation}
\left\langle V_{\alpha}(x)\right\rangle =\frac{U(\alpha|\mu_{B})}{\left|
z-\bar{z}\right|  ^{2\Delta_{\alpha}}} \label{onepoint}%
\end{equation}
In fig.\ref{fig1}a it is drawn however as the one-point function in the unit disk.


2. Second, one needs the boundary two-point function
\begin{equation}
\left\langle B_{\beta}^{\mu_{1}\mu_{2}}(x)B_{\beta}^{\mu_{2}\mu_{1}%
}(0)\right\rangle =\frac{d(\beta|\mu_{1},\mu_{2})}{\left|  x\right|
^{2\Delta_{\beta}}} \label{b2point}%
\end{equation}
which in general depends on two boundary cosmological constants $\mu_{1}$and
$\mu_{2}$ (see fig.\ref{fig1}b).

3. The bulk-boundary structure constant, which determines the fusion of a bulk
operator $V_{\alpha}$ with the boundary resulting in the boundary operator
$B_{\beta}^{\mu_{B}\mu_{B}}$. This is basically the same as the bulk-boundary
two-point function (fig.\ref{fig1}c)
\begin{equation}
\left\langle V_{\alpha}(z)B_{\beta}(x)\right\rangle =\frac{R(\alpha,\beta
|\mu_{B})}{\left|  z-\bar{z}\right|  ^{2\Delta_{\alpha}-\Delta_{\beta}}\left|
z-x\right|  ^{2\Delta_{\beta}}} \label{bbound}%
\end{equation}
In fact the one-point function (\ref{onepoint}) is a particular case of this
quantity with $\beta=0$ so that its introduction, however convenient, is redundant.

4. Finally, there is a boundary three-point function
\begin{equation}
\left\langle B_{\beta_{1}}^{\mu_{2}\mu_{3}}(x_{1})B_{\beta_{2}}^{\mu_{3}%
\mu_{1}}(x_{2})B_{\beta_{3}}^{\mu_{1}\mu_{2}}(x_{3})\right\rangle
=\frac{c(\beta_{1},\beta_{2},\beta_{3}|\mu_{1},\mu_{2},\mu_{3})}{\left|
x_{12}\right|  ^{\Delta_{1}+\Delta_{2}-\Delta_{3}}\left|  x_{23}\right|
^{\Delta_{2}+\Delta_{3}-\Delta_{1}}\left|  x_{31}\right|  ^{\Delta_{3}%
+\Delta_{1}-\Delta_{2}}} \label{b3point}%
\end{equation}
which in fact depends now on three different boundary parameters, $\mu_{1}$,
$\mu_{2}$ and $\mu_{3}$ related to the corresponding sides of the triangle as
shown in fig.\ref{fig1}d.

These three basic boundary structure constants, together with the bulk
structure constants, allow in principle to write down an intermediate state
expansions for any multipoint function. An instructive example is the bulk
two-point function
\begin{equation}
G_{\alpha_{1}\alpha_{2}}(z_{1},z_{2})=\left\langle V_{\alpha_{1}}%
(z_{1})V_{\alpha_{2}}(z_{2})\right\rangle \label{bulk2}%
\end{equation}
Joining these two operators together with the bulk structure constant we can
reduce this quantity to the one-point bulk function and write down the
following expansion
\begin{equation}
G_{\alpha_{1}\alpha_{2}}=\frac{\left|  z_{2}-\bar{z}_{2}\right|  ^{2\Delta
_{1}-2\Delta_{2}}}{\left|  z_{1}-\bar{z}_{2}\right|  ^{4\Delta_{1}}}\int
\frac{dP}{\pi}C(\alpha_{1},\alpha_{2},Q/2+iP)\,\,U(Q/2-iP){\mathcal{F}}\left(
\begin{array}
[c]{cc}%
\alpha_{1} & \alpha_{1}\\
\alpha_{2} & \alpha_{2}%
\end{array}
\left|  P\right|  \eta\right)  \label{bulk4}%
\end{equation}
where
\begin{equation}
\eta=\frac{(z_{1}-z_{2})(\bar{z}_{1}-\bar{z}_{2})}{(z_{1}-\bar{z}_{2})(\bar
{z}_{1}-z_{2})}\label{crossratio}%
\end{equation}
is the projective invariant of the four points $z_{1}$, $z_{2}$, $\bar{z}_{1}$
and $\bar{z}_{2}$ and $\mathcal{F}$ is the same four-point conformal block as
enters the expansion of the four-point bulk function, see
(\ref{fourpoint},\ref{block}). Notice that while in that case it entered in a
sesquilinear combination (\ref{fourpoint}), here it appears linearly (J.Cardy
\cite{Cardy1}). Expansion (\ref{bulk4}) is suitable appropriate if the bulk
operators are close to each other, i.e., $\eta\rightarrow0$. Another
representation is suitable the limit $\eta\rightarrow1$ where the points
$z_{1}$ and $z_{2}$ approach boundary and the bulk operators can be expanded
in the boundary ones. This gives
\begin{equation}
G_{\alpha_{1}\alpha_{2}}=\frac{\left|  z_{2}-\bar{z}_{2}\right|  ^{2\Delta
_{1}-2\Delta_{2}}}{\left|  z_{1}-\bar{z}_{2}\right|  ^{4\Delta_{1}}}\int
\frac{dP}{\pi}R(\alpha_{1},Q/2+iP)R(\alpha_{2},Q/2-iP)\,{\mathcal{F}}\left(
\begin{array}
[c]{cc}%
\alpha_{1} & \alpha_{2}\\
\alpha_{1} & \alpha_{2}%
\end{array}
\left|  P\right|  1-\eta\right)  \label{bound4}%
\end{equation}
Equating these two expressions we see that the basic boundary quantities also
must satisfy some bootstrap relations analogous to that in the bulk case. It
is interesting to note, that there is another application of this relation.
The conformal block itself, although being completely determined by the
conformal symmetry, is it fact a complicated function which is not in general
known explicitly. On the other hand it is important since it explicitely
enters the conformal bootstrap equations \cite{BPZ}. Besides, one might expect
that it encodes some information about the structure of the representations of
the conformal symmetry. In particular conformal block must satisfy the
following cross-relation
\begin{equation}
{\mathcal{F}}\left(
\begin{array}
[c]{cc}%
\alpha_{1} & \alpha_{3}\\
\alpha_{2} & \alpha_{4}%
\end{array}
\left|  P\right|  \eta\right)  =\int\frac{dP^{\prime}}{2\pi}K\left(
\begin{array}
[c]{cc}%
\alpha_{1} & \alpha_{3}\\
\alpha_{2} & \alpha_{4}%
\end{array}
\mid P,P^{\prime}\right)  \,{\mathcal{F}}\left(
\begin{array}
[c]{cc}%
\alpha_{1} & \alpha_{2}\\
\alpha_{3} & \alpha_{4}%
\end{array}
\left|  P^{\prime}\right|  1-\eta\right)  \label{crossmatrix}%
\end{equation}
with some cross-matrix $K$ which determines the monodromy properties of the
conformal block. Suppose now we've managed to find the basic quantities of the
boundary Liouville problem, in particular the one-point function $U(\alpha)$
and the bulk-boundary structure constant $R(\alpha,\beta)$. Then the crossing
relation becomes a linear equation for the cross-matrix of the symmetric
(i.e., $\alpha_{3}=\alpha_{1}$ and $\alpha_{4}=\alpha_{2}$) conformal block,
from where this matrix can be figured out\footnote{In a recent paper
\cite{PonsotTeschner} an explicit expression for this matrix has been proposed
on the basis of completely different approach}.

\subsection{Bulk one-point function}

We start with the calculation of the bulk one point function $U(\alpha|\mu
_{B})$. For this we apply the degenerate operator insertion, like above for
the bulk two-point function. Consider the auxiliary bulk two-point function
with the additional degenerate bulk field $V_{-b/2}(z)$%
\begin{equation}
G_{\alpha,-b/2}(x,z)=\left\langle V_{\alpha}(x)V_{-b/2}(z)\right\rangle
\label{auxiliary}%
\end{equation}
Apply first the OPE at $z\rightarrow x$ where the degenerate operator
$V_{-b/2}$ generates only two primary fields so that
\begin{equation}
G_{\alpha,-b/2}=C_{+}\,(\alpha)U(\alpha-b/2){\mathcal{G}}_{+}(x,z)+C_{-}%
(\alpha)\,\,U(\alpha+b/2){\mathcal{G}}_{-}(x,z) \label{special3}%
\end{equation}
where $C_{\pm}(\alpha)$ are the special structure constants as given by the
screening integrals and ${\mathcal{G}}_{\pm}(x,z)$ are expressed through the
special conformal blocks ${\mathcal{F}}_{\pm}(x,z)$ related to these special
values of parameters
\begin{equation}
{\mathcal{G}}_{\pm}(x,z)=\frac{\left|  x-\bar x\right|  ^{2\Delta_{\alpha
}-2\Delta_{-b/2}}}{\left|  z-\bar x\right|  ^{4\Delta_{\alpha}}}%
{\mathcal{F}}_{\pm}(\eta) \label{gblocks}%
\end{equation}
where $\Delta_{-b/2}=-1/2-3b^{2}/4$ and
\begin{equation}
\eta=\frac{(z-x)(\bar z-\bar x)}{(z-\bar x)(\bar z-x)} \label{fourinv}%
\end{equation}
In fact $V_{-b/2}$ satisfies the second order differential equation. Therefore
these special conformal blocks are solution to a second order linear
differential equation and can be expressed in terms of the hypergeometric
functions
\begin{align}
{\mathcal{F}}_{+}(\eta)=  &  {\mathcal{F}}\left(
\begin{array}
[c]{cc}%
-b/2 & -b/2\\
\alpha & \alpha
\end{array}
|\alpha-\frac b2|\eta\right)  =\nonumber\\
&  \ \eta^{\alpha b}(1-\eta)^{-b^{2}/2}F(2\alpha b-1-2b^{2},-b^{2},2\alpha
b-b^{2},\eta)\label{specblocks}\\
{\mathcal{F}}_{-}(\eta)=  &  {\mathcal{F}}\left(
\begin{array}
[c]{cc}%
-b/2 & -b/2\\
\alpha & \alpha
\end{array}
|\alpha+\frac b2|\eta\right)  =\nonumber\\
&  \ \eta^{1+b^{2}-\alpha b}(1-\eta)^{-b^{2}/2}F(-b^{2},1-2\alpha
b,2+b^{2}-2\alpha b,\eta)\nonumber
\end{align}
This is a particular case of more general conformal block with a degenerate
operator $V_{-b/2}$%
\begin{align}
{\mathcal{F}}\left(
\begin{array}
[c]{cc}%
-b/2 & \alpha_{3}\\
\alpha_{2} & \alpha_{4}%
\end{array}
|\alpha_{2}-\frac b2|\eta\right)   &  =\eta^{b\alpha_{2}}(1-\eta)^{\alpha
_{3}b}F(A,B,C,\eta)\label{dblocks}\\
{\mathcal{F}}\left(
\begin{array}
[c]{cc}%
-b/2 & \alpha_{3}\\
\alpha_{2} & \alpha_{4}%
\end{array}
|\alpha_{2}+\frac b2|\eta\right)   &  =\eta^{1+b^{2}-b\alpha_{2}}%
(1-\eta)^{b\alpha_{3}}F(A-C+1,B-C+1,2-C,\eta)\nonumber
\end{align}
where
\begin{align}
A  &  =-1+b(\alpha_{2}+\alpha_{3}+\alpha_{4}-3b/2)\nonumber\\
B  &  =b(\alpha_{2}+\alpha_{3}-\alpha_{4}-b/2)\label{param}\\
C  &  =b(2\alpha_{2}-b)\nonumber
\end{align}

Now, as both operators approach the boundary, they are expanded in the
boundary operators. It turns out that the degenerate bulk operator $V_{-b/2}$
near the boundary gives rise to only two primary boundary families $B_{0}$ and
$B_{-b}$. The simplest thing is to find the contribution of $B_{0}=I$. The
fusion of $V_{\alpha}$ to the unity boundary operator is described by the
quantity $R(\alpha,0)=v(\alpha)$ while the fusion of the field $V_{-b/2}$
($V_{-b/2}\rightarrow$ boundary) is described by a special bulk-boundary
structure constant $R(-b/2,Q).$ It can be computed as the following boundary
screening integral with one insertion of the boundary interaction $-\mu
_{B}\int B_{b}(x)dx$%
\begin{align}
R(-b/2,Q)  &  =-2^{2\Delta_{12}}\mu_{B}\int\left\langle V_{-b/2}%
(i)B_{b}(x)B_{Q}(\infty)\right\rangle dx\label{boundint}\\
\  &  =-\frac{2\pi\mu_{B}\Gamma(-1-2b^{2})}{\Gamma^{2}(-b^{2})}\nonumber
\end{align}
Comparing this with the behavior predicted by the bulk expansion
(\ref{special3}) we find the following functional equation for the one-point
function
\begin{equation}
-\frac{2\pi\mu_{B}}{\Gamma(-b^{2})}U(\alpha)=\frac{\Gamma(-b^{2}+2b\alpha
)}{\Gamma(-1-2b^{2}+2b\alpha)}U(\alpha-b/2)-\frac{\pi\mu\Gamma(-1-b^{2}%
+2b\alpha)}{\gamma(-b^{2})\Gamma(2b\alpha)}U(\alpha+b/2) \label{vrelation}%
\end{equation}
(in the last term we used the bulk special structure constant $C_{-}(\alpha) $
from(\ref{Cplus})). Equation is solved by the following simple expression
\begin{equation}
U(\alpha)=\frac2b\left(  \pi\mu\gamma(b^{2})\right)  ^{(Q-2\alpha)/2b}%
\Gamma(2b\alpha-b^{2})\Gamma\left(  \frac{2\alpha}b-\frac1{b^{2}}-1\right)
\cosh(2\alpha-Q)\pi s \label{vexplicit}%
\end{equation}
where the parameter $s$ is related to the scale invariant ratio of the
cosmological constants
\begin{equation}
\cosh^{2}\pi bs=\frac{\mu_{B}^{2}}\mu\sin\pi b^{2} \label{mumub}%
\end{equation}
Also this expression satisfies the dual functional equation provided the dual
bulk cosmological constant $\tilde\mu$ is related to $\mu$ as before in
(\ref{mutwiddle}) while the parameter $s$ is self-dual, i.e. the dual boundary
cosmological constant $\tilde\mu_{B}$ is defined as follows
\begin{equation}
\cosh^{2}\frac{\pi s}b=\frac{\tilde\mu_{B}^{2}}{\tilde\mu}\sin\frac\pi{b^{2}}
\label{mumubdual}%
\end{equation}

It is remarkable enough that the expression (\ref{vexplicit}) automatically
satisfies the ``reflection relation'' \cite{AAl} for the operator $V_{\alpha}
$%
\begin{equation}
U(\alpha)=U(Q-\alpha)D(\alpha) \label{vreflection}%
\end{equation}
with the bulk Liouville two-point function (\ref{Dexplicit}). If $\alpha$
corresponds to a physical state, i.e., $\alpha=Q/2+iP$ with $P$ real,
expression (\ref{vexplicit}) reads
\begin{equation}
U(P)=\left(  \pi\mu\gamma(b^{2})\right)  ^{-iP/b}\Gamma\left(  1+2ibP\right)
\Gamma\left(  1+2iP/b\right)  \frac{\cos\left(  2\pi sP\right)  }{iP}
\label{lexplicit}%
\end{equation}
This quantity is interpreted as the matrix element between a primary physical
state $\left|  v_{P}\right\rangle $ from (\ref{physspace}) and the boundary
state $\left\langle B_{s}\right|  $ created by the boundary action
(\ref{bound})
\begin{equation}
U(P)=\left\langle B_{s}|v_{P}\right\rangle \label{bstate}%
\end{equation}
It is natural that this matrix element satisfies the reflection relation
\begin{equation}
U(P)=S(P)U(-P) \label{lrefl}%
\end{equation}
with the Liouville reflection amplitude \cite{AAl}
\begin{equation}
S_{\mathrm{L}}(P)=-\left(  \pi\mu\gamma(b^{2})\right)  ^{-2iP/b}\frac
{\Gamma(1+2iP/b)\Gamma(1+2iPb)}{\Gamma(1-2iP/b)\Gamma(1-2iPb)}
\label{lreflection}%
\end{equation}

Of course the functional relation does not fix the overall constant so that it
can be multiplied by any (self-dual) factor $U_{0}(b)=U_{0}(1/b)$. In
(\ref{vexplicit}) this factor is chosen in the way that all the residues in
the ``on-mass-shell'' poles at $2\alpha=Q-nb$, with $n=1,2,3,\ldots$ are
\emph{equal precisely }to the corresponding perturbative integrals appearing
in expansions in $\mu$ and $\mu_{B}$, i.e.
\begin{align}
{{\operatorname*{res}}_{2\alpha=Q-nb}}  &  \frac{U(\alpha)}{2^{2\Delta
_{\alpha}}}=\sum\limits_{k=0}^{[n/2]}\frac{(-\mu)^{k}(-\mu_{B})^{n-2k}%
}{k!(n-2k)!}\times\label{residue}\\
&  \ \int\limits_{\mathrm{Im}z_{i}>0}d^{2}z_{1}\ldots d^{2}z_{k}%
\int\limits_{-\infty}^{\infty}dx_{1}\ldots dx_{n-2k}\left\langle
V_{Q-nb}(i)V_{b}(z_{1})\ldots V_{b}(z_{k})B_{b}(x_{1})\ldots B_{b}%
(x_{n-2k})\right\rangle _{0}\nonumber
\end{align}
where $\left\langle \cdots\right\rangle _{0}$ is the correlation function
w.r.t the upper half plane free field with $\mu=\mu_{B}=0$ i.e., with free
boundary conditions. Explicitely
\begin{equation}
\left\langle V_{\alpha_{1}}(z_{1})\ldots V_{\alpha_{n}}(z_{n})B_{\beta_{1}%
}(x_{1})\ldots B_{\beta_{m}}(x_{m})\right\rangle _{0}=\frac{\prod
\limits_{i=1}^{n}\left|  z_{i}-\bar z_{i}\right|  ^{-2\alpha_{i}^{2}}%
\prod\limits_{i,j}\left|  z_{i}-x_{j}\right|  ^{-4\alpha_{i}\beta_{j}}}%
{\prod\limits_{i>j}^{m}\left|  x_{i}-x_{j}\right|  ^{2\beta_{i}\beta_{j}}%
\prod\limits_{i>j}^{n}\left|  (z_{i}-z_{j})(z_{i}-\bar z_{j})\right|
^{4\alpha_{i}\alpha_{j}}} \label{free}%
\end{equation}
In particular, the pure boundary perturbations in $\mu_{B}$ reproduce the
Dyson integrals over a unit circle
\begin{equation}
\oint\frac{\prod_{i}^{n}du_{i}}{\prod_{i>j}^{n}\left|  u_{i}-u_{j}\right|
^{2b^{2}}}=\left(  \frac{2\pi}{\Gamma(1-b^{2})}\right)  ^{n}\Gamma(1-nb^{2})
\label{dyson}%
\end{equation}

Several remarks are in order in connection with the expression
(\ref{vexplicit}) presented.

\textbf{1. Semiclassical tests. }Consider the limit $b\rightarrow0$ while $P $
in eq.(\ref{lexplicit}) is of order $b$ and $s$ is of order $b^{-1}$. In this
limit the minisuperspace approximation is expected to work. Take the geometry
of semi-infinite cylinder of circumference $2\pi$ and consider the states on
the circle. In the minisuperspace approximation one takes into account the
dynamics of the zero mode
\begin{equation}
\phi_{0}=\frac1{2\pi}\int_{0}^{2\pi}\phi(x)dx \label{zeromode}%
\end{equation}
neglecting completely all the oscillator modes of field $\phi(x)$. The primary
state $\left|  v_{P}\right\rangle $ is represented now by the wave function
\begin{equation}
\psi_{P}(\phi_{0})=\frac{2\left(  \pi\mu/b^{2}\right)  ^{-iP/b}}{\Gamma\left(
-2iP/b\right)  }K_{2iP/b}\left(  2\sqrt{\pi\mu/b^{2}}e^{b\phi_{0}}\right)
\label{psiclassic}%
\end{equation}
($K_{v}(z)$ is the modified Bessel function) which satisfies the
minisuperspace Shr\"odinger equation
\begin{equation}
\left(  -\frac12\frac{d^{2}}{d\phi_{0}^{2}}+2\pi\mu e^{2b\phi_{0}}%
-2P^{2}\right)  \psi_{P}(\phi_{0})=0 \label{shrodinger}%
\end{equation}
and has the following asymptotic at $\phi_{0}\rightarrow-\infty$%
\begin{equation}
\psi_{P}(\phi_{0})=e^{2iP\phi_{0}}+S_{\text{\textrm{L}}}^{\mathrm{(cl)}%
}(P)e^{-2iP\phi_{0}} \label{waves}%
\end{equation}
where $S_{\text{\textrm{L}}}^{\mathrm{(cl)}}(P)$ is the classical limit of the
Liouville reflection amplitude (\ref{lreflection}). Also it meets the
normalization (\ref{norm})
\begin{equation}
\int_{-\infty}^{\infty}\psi_{P^{\prime}}^{*}(\phi_{0})\psi_{P}(\phi_{0}%
)=\pi\delta(P-P^{\prime}) \label{clnorm}%
\end{equation}
In the approximation under consideration the boundary state $\left\langle
B_{s}\right|  $ wave function is simply related to the boundary Lagrangian
\begin{equation}
\Psi_{B_{s}}(\phi_{0})=\exp(-2\pi\mu_{B}e^{b\phi_{0}}) \label{bwf}%
\end{equation}
The matrix element $\left\langle B_{s}|v_{P}\right\rangle $ can be carried out
explicitely
\begin{equation}
\int_{-\infty}^{\infty}\Psi_{B_{s}}(\phi_{0})\psi_{P}(\phi_{0})d\phi_{0}%
=\frac2b\left(  \pi\mu/b^{2}\right)  ^{-iP/b}\Gamma\left(  2iP/b\right)
\cos(2\pi Ps) \label{element}%
\end{equation}
and agrees precisely with the corresponding limit of (\ref{lexplicit}). Note,
that this calculation is sensitive to the prefactor in eq.(\ref{vexplicit}) and
confirms our choice $U_{0}(b)=1$, at least in the limit $b\rightarrow0$.

\textbf{2. Boundary length distribution.} From the point of view of 2D gravity
one can interprete the quantity
\begin{equation}
l=\int_{\partial\Gamma}\exp b\phi\label{length}%
\end{equation}
as the lenth of the boundary. Let $W_{\alpha}(l)$ be the boundary length
distribution for the fluctuating disk with the bulk cosmological constant
$\mu$ and an insertion of the operator $V_{\alpha}$ somewhere inside the disk.
Then
\begin{equation}
U(\alpha|\mu_{B})=\int_{0}^{\infty}\frac{dl}{l}e^{-\mu_{B}l}W_{\alpha
}(l)\label{ww}%
\end{equation}
Form the result (\ref{vexplicit}) one finds explicitely
\begin{equation}
W_{\alpha}(l)=\frac{2}{b}\left(  \pi\mu\gamma(b^{2})\right)  ^{(Q-2\alpha
)/2b}\frac{\Gamma(2\alpha b-b^{2})}{\Gamma(1+1/b^{2}-2\alpha/b)}%
K_{(Q-2\alpha)/b}(\kappa l)\label{distribution}%
\end{equation}
where
\begin{equation}
\kappa^{2}=\frac{\mu}{\sin\pi b^{2}}\label{kappa}%
\end{equation}
Compare (\ref{distribution}) with the minisuperspace distribution
(\ref{psiclassic}). This result implies that the Shr\"{o}dinger equation
(\ref{shrodinger}) in the logarithm of the scale $\phi_{0}=b^{-1}\log(l/2\pi)$
(which is called sometimes the Wheeler-deWitt equation) 
does not hold only in semiclassical limit but is in fact exact with a suitable
renormalizations of constants (see in this relation the paper
\cite{Moore} where this equation first appeared in the context of the
Liouville field theory, and also \cite{Kostov1, Kostov2} where similar
expressions are obtained in the framework of random surface models).

Let us present also the double distribution in the length (\ref{length}) and
area $A$ defined as
\begin{equation}
W_{\alpha}(l,\mu)=\int_{0}^{\infty}\frac{dA}Ae^{-\mu A}Z_{\alpha}(A,l)
\label{area}%
\end{equation}
It is given by a rather simple expression
\begin{equation}
Z_{\alpha}(A,l)=\frac1b\frac{\Gamma(2\alpha b-b^{2})}{\Gamma(1+1/b^{2}%
-2\alpha/b)}\left(  \frac{l\Gamma(b^{2})}{2A}\right)  ^{(Q-2\alpha)/b}%
\exp\left(  -\frac{l^{2}}{4A\sin\pi b^{2}}\right)  \label{aread}%
\end{equation}

\textbf{3. ``Heavy'' }$\alpha$ \textbf{semiclassics. }Consider again the limit
$b\rightarrow0$ but with large value of $\alpha=\eta/b$ not nessesserily close
to $Q/2$. Exact expression (\ref{aread}) gives in this limit
\begin{equation}
Z_{\eta/b}(A,l)=\frac{l\Gamma(2\eta)e^{(\eta-1/2)C}}{2b^{2}A\sqrt{2\pi
(1-2\eta}}\exp\left(  -\frac1{b^{2}}S_{\eta}(A,l)\right)  \label{semicl2}%
\end{equation}
where $C$ is the Euler's constant and
\begin{equation}
S_{\eta}(A,l)=\frac{l^{2}}{4\pi A}+(1-2\eta)\left(  \log\left(  2A/l\right)
+\log(1-2\eta)-1\right)  \label{claction}%
\end{equation}

On the other hand the corresponding classical solution with the area $A$ and
the boundary lenght $l$ reads for the classical field $\varphi=2b\phi$ (we
imply here the geometry of the disk $\left|  z\right|  \leq1$ with the unit
circle as the boundary)
\begin{equation}
e^{\varphi(z)}=\frac{l^{2}(a^{-1}-a)^{2}}{\left[  2\pi(a^{-1}(z\bar z)^{\eta
}-a(z\bar z)^{1-\eta})\right]  ^{2}} \label{cliouv}%
\end{equation}
where $a$ is related to the area as follows
\begin{equation}
a^{2}=1-\frac{4\pi A}{l^{2}}(1-2\eta) \label{aparam}%
\end{equation}
(we imply here that $\eta\leq1/2$ and $l^{2}>4\pi A(1-2\eta)$ so that a real
classical solution exists). The classical Liouville action for this solution
is readily carried out
\begin{equation}
S_{\mathrm{cl}}(A,l)=\frac{l^{2}}{4\pi A}+(1-2\eta)\left(  \log\left(
2A/l\right)  +\log(1-2\eta)-1\right)
\label{sdisk}%
\end{equation}
and coincides with (\ref{claction}). In principle it might be possible to
check the prefactor in (\ref{semicl2}) performing the one-loop correction.
This is not yet done however.

\textbf{4.``Light'' }$\alpha$ \textbf{semiclassics. }Direct semiclassical
calculation of the one-point function (\ref{vexplicit}) is possible also in
the case $\alpha=b\sigma$, with $b\rightarrow0$ and $\sigma$ fixed. In
particular, one can calculate the semiclassical approximation to the function
(\ref{aread}) by taking the saddle-point contribution to the corresponding
functional integral over $\phi$ with fixed area $A$ and boundary length $l$.
In the present case $\alpha\sim b$ the exponential insertion does not affect
the saddle-point configurations. The nature of these classical solutions
depends on the relative value of $A$ and $l^{2}$. Here we consider 
explicitely only
the negative-curvature situation $4\pi A<l^{2}$, in which case the classical
configurations form an orbit under the action of $SL(2,R)$. To be specific we
adopt the upper half-plane geometry with the boundary at the real axis. Then
generic classical solution $\phi_{G}$ is obtained from the ``standard''
solution
\begin{equation}
e^{2b\phi_{I}}(z,{\bar{z}})\,dz\,d\bar{z}=\frac{\rho^{2}\,dz\,d\bar{z}%
}{((z+i)(\bar{z}-i)-\nu^{2})^{2}}\, \label{stsolution}%
\end{equation}
by $SL(2,R)$ transformation
\begin{equation}
z\rightarrow G\ast z=\frac{az+b}{cz+d},\qquad G=\left(
\genfrac{}{}{0pt}{}{a\ b}{c\ d}%
\right)  \in SL(2,R)\, \label{sl2r}%
\end{equation}
here
\begin{equation}
\rho=\frac{2lA}{l^{2}-2\pi A}\,,\qquad\nu=\frac{l\,\sqrt{l^{2}-4\pi A}}%
{l^{2}-2\pi A}\, \label{rhonu}%
\end{equation}
The semiclassical approximation to the expectation value (\ref{aread}) is then
evaluated as an integral over this manifold of classical configurations, i.e.
\begin{equation}
\left\langle e^{2b\sigma\phi}\right\rangle _{A,l}=N\,e^{-\frac{1}{b^{2}}
\,S_{\mathrm{cl}}}\,\int d\mu(G)\,%
\left(  e^{2b\phi_{G}}(z,{\bar{z}})\right)  ^{\sigma}
\label{fintegral}%
\end{equation}
where $S_{\mathrm{cl}}$
is the classical action (\ref{sdisk}) with $\eta=0$, $d\mu(G)$ stands for the 
$SL(2,R)$ invariant
integration mesure and the factor $N$ combines the determinant of zero modes
and the contributions of positive modes to the gaussian integral around given
classical solution. It is important to note that while $N$ can very well
depend on $A/l^{2}$, it carries no dependence on $\sigma$, i.e. all the
$\sigma$ dependence of the one-point function in this approximation comes from
the integral in (\ref{fintegral}).

The integrand in (\ref{fintegral}) can be simplified by a shift of the
integration variable $G\rightarrow G_{z}\,G$, where $G_{z}$ is any fixed
($z$-dependent) $SL(2,R)$ transformation which maps the point $z$ to the point
$i$ in the upper half-plane; this gives for the integral in (\ref{fintegral})
\begin{equation}
\left(  \frac{2i\,\rho}{z-\bar{z}}\right)  ^{2\sigma}\,\int\,d\mu
(G)\,((a^{2}+b^{2}+c^{2}+d^{2}+2)-\nu^{2}(c^{2}+d^{2}))^{-2\sigma}.
\label{fint2}%
\end{equation}
To evaluate this integral one can introduce the following coordinates on the
group manifold of $SL(2,R)$,
\begin{equation}
a/d=i\,\frac{2y\bar{y}-x(y+\bar{y})}{y-\bar{y}}\,;\qquad b/d=x\,;\qquad
c/d=i\,{\frac{y+\bar{y}-2x}{y-\bar{y}}}\,;\qquad d=\frac{1}{\sqrt
{\Im\mathrm{m}\ y}\,|y-x|} \label{xy}%
\end{equation}
where $x$ is real and $y$ and $\bar{y}$ are complex conjugate with
$\Im\mathrm{m\;}y>0$. The invariant mesure takes the form
\begin{equation}
d\mu(G)=\frac{2i\,dx\,d^{2}y}{(y-\bar{y})\,|y-x|^{2}}\, \label{xymeasure}%
\end{equation}
and the integral in (\ref{fint2}) can be written as
\begin{equation}
\int\left(  \frac{y-\bar{y}}{2i}\right)  ^{2\sigma-1}\,((y+i)(\bar{y}%
-i)-\nu^{2})^{-2\sigma}\frac{dx\,d^{2}y}{|y-x|^{2}}\, \label{fint3}%
\end{equation}
This integral is readily evaluated and one obtains for (\ref{fintegral})
\begin{equation}
\tilde{N}\,\frac{1}{2\sigma-1}\,\left(  \frac{i}{z-\bar{z}}\,\frac{2A}%
{l}\right)  ^{2\sigma}\,e^{-\frac{1}{{\,}b^{2}}\,S_{\mathrm{cl}}}%
\label{fint4}%
\end{equation}
where the factor $\tilde{N}=\pi\,l^{2}\,N/A$ does not depend on $\sigma$; as
is mentioned above its determination requires analysis of the fluctuations
around the classical configurations which we did not perform. The $\sigma
$-dependent part in (\ref{fint4}) agrees with $b\rightarrow0$ limit of
(\ref{aread}).

\textbf{5. Boundary state.} Once the function $U(P)$ is constructed, the
boundary state $\left\langle B_{s}\right|  $ can be written down explicitely
\begin{equation}
\left\langle B_{s}\right|  =\int_{0}^{\infty}\frac{dP}\pi U(P)\left\langle
P\right|  \label{state}%
\end{equation}
where the so called Ishibashi states \cite{Ishibashi}
\begin{equation}
\left\langle P\right|  =\left\langle v_{P}\right|  \left(  1+\frac{L_{1}\bar
L_{1}}{2P^{2}+Q^{2}/2}+\ldots\right)
\end{equation}
are designed in the way to match the conformal invariance of the boundary.
Since the combination $U(P)\left\langle P\right|  $ is invariant w.r.t. the
reflection $P\rightarrow-P$ one can extend formally the integral (\ref{state})
to the negative values of $P$ and write
\begin{equation}
\left\langle B_{s}\right|  =\int_{-\infty}^{\infty}\frac{dP}{2\pi}e^{2i\pi
Ps}u(P)\left\langle P\right|  \label{infinite}%
\end{equation}
where
\begin{equation}
u(P)=\frac1{iP}\left(  \pi\mu\gamma(b^{2})\right)  ^{-iP/b}\Gamma\left(
1+2ibP\right)  \Gamma\left(  1+2iP/b\right)
\end{equation}
It is natural to call $u(P)$ the boundary state wave function. Note that the
state $\left\langle P\right|  $, although consistent with the conformal
invariance, does not correspond to any conformal boundary state, i.e., to a
state created by a local conformally invariant boundary condition. However, it
can be constructed as a linear combination of boundary states. In view of
eq.(\ref{infinite}) we can write down
\begin{equation}
\frac{u(P)}{2\pi}\left\langle P\right|  =\int_{-\infty}^{\infty}e^{-2i\pi
Ps}\left\langle B_{s}\right|  ds \label{inversion}%
\end{equation}
This equation allows to single out a conformally invariant state containing
only one primary state $\left\langle v_{P}\right|  $ and its descendents. In
finite dimensional situation of rational conformal field theories this trick
has been friquently used by J.Cardy \cite{Cardy2}.

\section{Boundary two-point function}

In this section the boundary two-point function $d(\beta|\mu_{1},\mu_{2})$ of
(\ref{b2point}) will be derived. To this purpose we apply basically the same
Techner's tric which has been used in the first section to determine the bulk
structure constants. Considering the boundary operators $B_{\beta}(x)$ we find
that all the operators $B_{-nb/2}(x)$ (and also of course the dual fields
$B_{-n/2b})$ with $n=0,1,\ldots$are degenerate, i.e., count primary states
among their descendents. A complication here is that not all of these ``null
vectors'' nessesserily vanish, contrary to what happens in the bulk situation.
For example simplest non-trivial degenerate boundary operator $B_{-b/2}%
^{s_{1}s_{2}}$ (form now on we shall denote the exponential boundary operators
$B_{\beta}^{s_{1}s_{2}}=e_{s_{1}s_{2}}^{\beta\phi}$ instead of $B_{\beta}%
^{\mu_{1}\mu_{2}}$ having in mind the relation (\ref{mumub})) in general does
not satisfy the second order differential equation. This means that the
null-vector in the corresponding Virasoro representation is some non-vanishing
primary field and therefore the second order differential equation has some
non-zero terms in the right hand side. This effect can be already seen at the
classical level where the upper half-plane boundary Liouville problem is
reduced to the classical Liouville equation for the field $\varphi=2b\phi$%
\begin{equation}
\partial\bar\partial\varphi=2\pi\mu b^{2}e^{\varphi} \label{Lcl}%
\end{equation}
in the upper half-plane with the boundary condition
\begin{equation}
i(\partial-\bar\partial)\varphi=4\pi\mu_{B}b^{2}e^{\varphi/2} \label{bLcl}%
\end{equation}
at the real axis. The boundary value of the classical stress tensor can be
easily computed
\begin{equation}
T_{\mathrm{cl}}(x)=-\frac1{16}\varphi_{x}^{2}+\frac14\varphi_{xx}+\pi
b^{2}(\pi\mu_{B}^{2}b^{2}-\mu)e^{\varphi} \label{bclstress}%
\end{equation}
The boundary operator $B_{-b/2}^{s,s}$ in the classical limit reduces to the
boundary value of $\exp(-\varphi/4)$ for which we have
\begin{equation}
\left(  \frac{d^{2}}{dx^{2}}+T_{\mathrm{cl}}\right)  e_{s,s}^{-\varphi/4}=\pi
b^{2}(\pi\mu_{B}^{2}b^{2}-\mu)e_{s,s}^{3\varphi/4} \label{bdiff2}%
\end{equation}
In the right-hand side there is a primary Virasoro operator $\exp
(3\varphi/4)_{s,s}$ which has exactly the same dimension as the null-vector in
the corresponding degenerate representation. It is interesting to note that
there is a unique relation between the cosmological constants $\pi\mu_{B}%
^{2}b^{2}/\mu=1$ where the r.h.s vanishes and this operator satisfies
homogeneous linear differential equation. This effect holds on the quantum
level too: if the boundary and bulk cosmological constants are related as
\begin{equation}
1=\frac{2\mu_{B}^{2}}\mu\tan\frac{\pi b^{2}}2 \label{tg}%
\end{equation}
the second order differential equation holds for the boundary operator
$B_{-b/2}^{s,s}$ (see the remark in the concluding section).

Here we are interested in the general situation where this operator is of no
use since it does not always satisfy the second order differential equation.
It happens however that the next degenerate boundary operator $B_{-b}^{s,s}$
do satisfy the third-order differential equation when placed between identical
boundary conditions. Therefore it can be used in our calculations instead of
$B_{-b/2}$. As in the bulk, the differential equation predicts the following
truncated OPE of this operator with any exponential boundary primary
\begin{equation}
B_{-b}^{s,s}B_{\beta}^{s,s^{\prime}}=c_{+}[B_{\beta-b}]+c_{0}[B_{\beta}%
]+c_{-}[B_{\beta+b}] \label{special13}%
\end{equation}
where $c_{\sigma}(\beta)$ are again the special boundary structure constants,
which can be calculated as certain screening integrals. Considering again the
auxilary three-point boundary function with a $B_{-b}$ insertion one figures
out immediately that
\begin{equation}
\frac{d(\beta+b|\mu_{1},\mu_{2})}{d(\beta|\mu_{1},\mu_{2})}=c_{-}^{-1}(\beta)
\label{bshift}%
\end{equation}

\begin{figure}
[tbh]
\begin{center}
\includegraphics[
height=0.8847in,
width=5.0099in
]%
{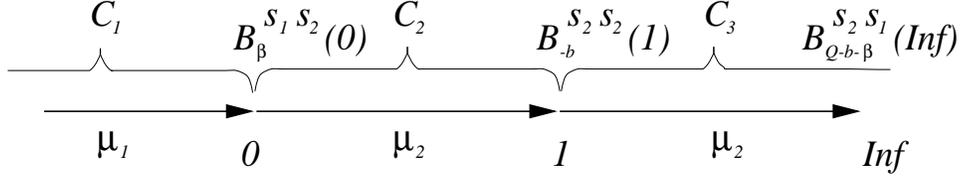}%
\caption{Contours of integration $C_{i}$ in (\ref{cint}) together with the
corresponding values of the boundary cosmological constants $\mu_{i}$.}%
\label{fig2}%
\end{center}
\end{figure}

The structure constant $c_{-}(\beta)$ can be evaluated as a combination of
scrining integrals. These are of two tipes: a volume screening by the bulk
Liouville interaction term $e^{2b\phi}$%
\begin{align}
c_{-}^{\mathrm{(v)}}  &  =-\mu\int\limits_{\mathrm{Im}z>0}d^{2}z\left\langle
e^{2b\phi(z)}B_{\beta}^{s_{1}s_{2}}(0)B_{-b}^{s_{2}s_{2}}(1)B_{Q-\beta
-b}^{s_{2}s_{1}}(\infty)\right\rangle \label{bulkint}\\
\  &  =\mu\int\limits_{\mathrm{Im}z>0}d^{2}z\frac{\left|  1-z\right|
^{4b^{2}}}{\left|  z\right|  ^{4b\beta}\left|  z-\bar{z}\right|  ^{2b^{2}}%
}\nonumber
\end{align}
and two boundary screenings $e^{b\phi}$ related to the boundary interaction
\begin{align}
c_{-}^{\mathrm{(b)}}  &  =\sum_{i,j}\frac{\mu_{i}\mu_{j}}{2}\int_{C_{i}}%
\int_{C_{j}}dx_{1}dx_{2}\left\langle e^{b\phi(x_{1})}e^{b\phi(x_{2})}B_{\beta
}^{s_{1}s_{2}}(0)B_{-b}^{s_{2}s_{2}}(1)B_{Q-\beta-b}^{s_{2}s_{1}}%
(\infty)\right\rangle \label{cint}\\
&  =\sum_{i,j}\frac{\mu_{i}\mu_{j}}{2}\int_{C_{i}}\int_{C_{j}}dx_{1}%
dx_{2}\frac{\left|  (1-x_{1})(1-x_{2})\right|  ^{2b^{2}}}{\left|  x_{1}%
-x_{2}\right|  ^{2b^{2}}\left|  x_{1}x_{2}\right|  ^{2b\beta}}\nonumber
\end{align}
where the contours $C_{i}$ are chosen as in fig.\ref{fig2} while $\mu_{i}$ are
the corresponding values of the boundary cosmological constant, as it is also
indicated in in the same figure. Both contributions can be carried out
explicitely and we have
\begin{align}
c_{-}(\beta)  &  =c_{-}^{\mathrm{(v)}}+c_{-}^{\mathrm{(b)}}\nonumber\\
&  =\left[  -\mu\sin^{2}(2\pi b\beta)+\sin\pi b^{2}\left(  \mu_{1}^{2}+\mu
_{2}^{2}-2\mu_{1}\mu_{2}\cos(2\pi b\beta)\right)  \right]  I_{0}%
(\beta)\label{cminus}\\
&  =4\mu\sin\left(  \pi b\frac{2\beta+i(s_{1}+s_{2})}{2}\right)  \sin\left(
\pi b\frac{2\beta-i(s_{1}+s_{2})}{2}\right)  \times\nonumber\\
&  \;\;\;\;\;\;\sin\left(  \pi b\frac{2\beta+i(s_{1}-s_{2})}{2}\right)
\sin\left(  \pi b\frac{2\beta-i(s_{1}-s_{2})}{2}\right)  I_{0}(\beta)\nonumber
\end{align}
where
\begin{equation}
I_{0}(\beta)=-\frac{\gamma(1+b^{2})}{\pi}\Gamma(1-2b\beta)\Gamma
(2b\beta-1)\Gamma(1-b^{2}-2b\beta)\Gamma(2b\beta-b^{2}-1) \label{I0}%
\end{equation}
and $s_{1}$ and $s_{2}$ are again related to $\mu_{1}$ and $\mu_{2}$ as in
eq.(\ref{mumub}).%


To construct a solution to the functional equation (\ref{bshift}) with
(\ref{cminus}) we need more special functions. First one is what is sometimes
called the $q$-gamma function. Here we denote it ${\mathbf{S}}(x)$. It is self
dual with repect to $b\rightarrow1/b$ and satisfies the following shift
relations
\begin{align}
{\mathbf{S}}(x+b)  &  =2\sin(\pi bx){\mathbf{S}}(x)\label{shift}\\
{\mathbf{S}}(x+1/b)  &  =2\sin(\pi x/b){\mathbf{S}}(x)\nonumber
\end{align}
It has zeroes at $x=Q+nb+m/b$ and poles at $x=-nb-m/b$ ( $m$ and $n$ are
non-negative integer numbers). In the strip $0<\mathrm{Re}$ $x<Q$ the
following integral representation is allowed
\begin{equation}
\log{\mathbf{S}}(x)=\int\limits_{0}^{\infty}\frac{dt}t\left[  \frac
{\sinh(Q-2x)t}{2\sinh(bt)\sinh(t/b)}-\frac{(Q/2-x)}t\right]  \label{intS}%
\end{equation}
With this definition it satisfies also the ``unitarity'' relation
\begin{equation}
{\mathbf{S}}(x){\mathbf{S}}(Q-x)=1 \label{unitS}%
\end{equation}
It is also convenient to introduce a self-dual entire function $\mathbf{G}(x)
$ which contains only zeroes at $x=-nb-m/b$, $m,n=0,1,2,\ldots$ and enjoes the
following shift relations
\begin{align}
{\mathbf{G}}(x+b)  &  =\frac{b^{1/2-bx}}{\sqrt{2\pi}}\Gamma(bx){\mathbf{G}}%
(x)\label{shiftG}\\
{\mathbf{G}}(x+1/b)  &  =\frac{b^{x/b-1/2}}{\sqrt{2\pi}}\Gamma(x/b)
{\mathbf{G}}%
(x)\nonumber
\end{align}
This function is ``elementary'' in the sense that both $\Upsilon(x)$ from
eq.(\ref{Upsilon}) and ${\mathbf{S}}(x)$ are simply expressed in 
${\mathbf{G}}(x)
$%
\begin{align}
\Upsilon(x)  &  ={\mathbf{G}}(x){\mathbf{G}}(Q-x)\nonumber\\
{\mathbf{S}}(x)  &  =\frac{{\mathbf{G}}(Q-x)}{{\mathbf{G}}(x)} \label{YSG}%
\end{align}
The integral representation which is valid for all $0<\mathrm{Re}$ $x$ reads
\begin{equation}
\log{\mathbf{G}}(x)=\int\limits_{0}^{\infty}\frac{dt}t\left[  \frac
{e^{-Qt/2}-e^{-xt}}{(1-e^{-bt})(1-e^{-t/b})}+\frac{(Q/2-x)^{2}}2e^{-t}%
+\frac{Q/2-x}t\right]  \label{intG}%
\end{equation}
With this function one can easily construct a solution to (\ref{bshift}).
\begin{align}
d(\beta|s_{1},s_{2})  &  =\frac{\left(  \pi\mu\gamma(b^{2})b^{2-2b^{2}%
}\right)  ^{(Q-2\beta)/2b}{\mathbf{G}}(Q-2\beta){\mathbf{G}}^{-1}(2\beta
-Q)}{{\mathbf{S}}(\beta+i(s_{1}+s_{2})/2){\mathbf{S}}(\beta-i(s_{1}+s_{2}%
)/2){\mathbf{S}}(\beta+i(s_{1}-s_{2})/2){\mathbf{S}}(\beta-i(s_{1}-s_{2}%
)/2)}\label{dexplicit}\\
&  =\left(  \pi\mu\gamma(b^{2})b^{2-2b^{2}}\right)  ^{(Q-2\beta)/2b}%
\frac{{\mathbf{G}}(Q-2\beta)}{{\mathbf{G}}(2\beta-Q)}\times\nonumber\\
&  \;\;\;\;\exp\left(  -%
{\displaystyle\int\limits_{-\infty}^{\infty}}
\frac{dt}t\left[  \frac{\sinh(Q-2\beta)t\cos s_{1}t\cos s_{2}t}{\sinh bt\sinh
t/b}-\frac{(Q-2\beta)}t\right]  \right) \nonumber
\end{align}
This solution satisfies also the ``dual-shift'' relation analogous to
(\ref{bshift}) so that (\ref{dexplicit}) is the unique self-dual solution to
(\ref{bshift}). It is of course possible to express the ratio of two
$\mathbf{G}$-functions in terms of $\mathbf{S}$-function times some ordinary
$\Gamma$-functions. We prefere to present $d(\beta|s_{1},s_{2})$ in the form
(\ref{dexplicit}) to make obvious the ``unitarity'' relation
\begin{equation}
d(\beta|s_{1},s_{2})d(Q-\beta|s_{1},s_{2})=1 \label{dunitarity}%
\end{equation}
Note, that an overall independent of $\beta$ constant which is allowed by
(\ref{bshift}) and its dual is completely fixed by (\ref{dunitarity}).

\section{Concluding remarks}

\begin{itemize}
\item  Eq.(\ref{tg}) together with the structure of singularities of the
two-point function (\ref{dexplicit}) drop a hint at the suggestion that the
level 2 degenerate boundary operator $B_{-b/2}^{s_{1}s_{2}}(x)$ has a
vanishing null vector if and only if $s_{1}-s_{2}=\pm ib$ or $s_{1}+s_{2}=\pm
ib$ (the second condition is requred by the symmetry of boundary conditions
w.r.t. $s\rightarrow-s$). Let us verify this suggestion on the three-point
function with one boundary field $B_{-b/2}^{s,s\pm ib}(x)$, i.e., consider
\begin{equation}
\left\langle B_{\beta}^{s_{2}s_{1}}(x_{1})B_{-b/2}^{s_{1}s_{1}\pm
ib}(x)B_{\beta+b/2}^{s_{1}\pm ib,s_{2}}(x_{2})\right\rangle
\end{equation}
Under our suggestion $B_{-b/2}^{s,s\pm ib}(x)$ satisfies a second order
differential equation in $x$ and therefore has special operator product
expansion with any $B_{\beta}$
\begin{equation}
B_{-b/2}^{s,s\pm ib}B_{\beta}^{s\pm ib,s^{\prime}}=c_{+}^{(\pm)}[B_{\beta
-b/2}^{s,s^{\prime}}]+c_{-}^{(\pm)}[B_{\beta+b/2}^{s,s^{\prime}}%
]\label{special112}%
\end{equation}
Then, exactly the same trick which led to eq.(\ref{bshift}) gives the
following shift relation
\begin{equation}
\frac{d(\beta|s,s^{\prime})}{d(\beta+b/2|s\pm ib,s^{\prime})}=c_{-}^{(\pm
)}(\beta)\label{halfshift}%
\end{equation}
As usual we adopt the structure constant with no screenings requred
$c_{+}^{(\pm)}=1$. The structure constant $c_{-}^{(\pm)}$ is given by the
integral
\begin{equation}
c_{-}^{(\pm)}(\beta)=\int_{-\infty}^{\infty}\left\langle e^{b\phi(x)}B_{\beta
}^{s_{2}s_{1}}(0)B_{-b/2}^{s_{1}s_{1}\pm ib}(1)B_{Q-\beta-b/2}^{s_{1}\pm
ib,s_{2}}(\infty)\right\rangle dx
\end{equation}
This integral is evaluated quite easily (unlike (\ref{bulkint}) or
(\ref{cint}))
\begin{align}
c_{-}^{(\pm)}(\beta) &  =\frac{\Gamma(1-2b\beta)\Gamma(2b\beta-b^{2}%
-1)\Gamma(1+b^{2})}{\pi}\times\\
&  \ \left[  \mu_{1}\sin\pi(b^{2}-2b\beta)+\mu_{1}^{(\pm)}\sin\pi(2b\beta
)-\mu_{2}\sin\pi b^{2}\right]  \nonumber
\end{align}
where $\mu_{1}^{(\pm)}$ is determined by the relation
\begin{equation}
\cosh\pi b(s_{1}\pm ib)=\mu_{1}^{(\pm)}\sqrt{\sin\pi b^{2}/\mu}%
\end{equation}
After some simple algebra we obtain
\begin{align}
c_{-}^{(\pm)}(\beta) &  =2\left(  -\frac{\mu}{\pi\gamma(-b^{2})}\right)
^{1/2}\Gamma(1-2b\beta)\Gamma(2b\beta-b^{2}-1)\times\\
&  \ \sin\pi b\left(  \beta\mp ib(s_{1}+s_{2})/2\right)  \sin\pi b^{2}\left(
\beta\mp ib(s_{1}-s_{2})/2\right)  \nonumber
\end{align}
It is easy to see that the two-point function (\ref{dexplicit}) satisfies both
relations (\ref{halfshift}). After this support one may suggest further that
any degenerate field $B_{-nb/2}^{s,s^{\prime}}$ has vanishing null-vector (and
therefore has truncated operator product expansions if $s-s^{\prime}=ibk$ or
$s+s^{\prime}=ibk$ with $k=-n/2,-n/2+1,-n/2+2,\ldots,n/2$, in close analogy
with the fusion rules for degenerate bulk fields.

\item  The boundary two-point function (\ref{dexplicit}) is readily applied as
the reflection coefficient in the reflection relations for the one-point
function of an exponetial boundary operator in the boundary sin-Gordon model.
The latter is defined by the following two-dimensional euclidean action
\begin{equation}
A_{\mathrm{bsG}}=\int\limits_{\Gamma}d^{2}x\,\left[  \frac{1}{{4\pi}%
}(\partial_{a}\phi)^{2}-2\mu\,\cos(2\beta\phi)\right]  -2\mu_{B}%
\int\limits_{\partial\Gamma}\cos(\beta(\phi-\phi_{0}))\label{bsing}%
\end{equation}
where the bulk part of the action is integrated over a half-plane $\Gamma$ so
that the boundary $\partial\Gamma$ is a strainght line. For the moment $\beta$
denotes the standard sin-Gordon coupling constant. Apart from it the boundary
model depends of three parameters $\mu$, $\mu_{B}$ and $\phi_{0}$
\cite{ghoshal}. The dimensional parameters $\mu$ and $\mu_{B}$ can be given a
precise meaning by specifying the normalisation of the composite fields they
couple to. As these operators are combinations of exponentials it suffices to
specify a normalisation for the exponential fields in the volume and at the
boundary. Here we adopt the conventional normalisation of these fields (see
e.g.\cite{bsg}) corresponding to the short distance asymptotics at $\left|
x-y\right|  \rightarrow0$%
\begin{align}
e^{2ia\phi}(x)e^{-2ia\phi}(y) &  =\frac{1}{\left|  x-y\right|  ^{4a^{2}}%
}+\ldots\;\text{\ \ \ \ \ \ \ \ for the volume fields}\nonumber\\
e^{ia\phi_{B}}(x)e^{-ia\phi_{B}}(y) &  =\frac{1}{\left|  x-y\right|  ^{2a^{2}%
}}+\ldots\;\;\;\;\;\;\;\text{for the boundary ones}%
\end{align}
Here we present only the result for the one-point function of the boundary
operator ${\mathcal{G}}_{B}(a)=\left\langle \exp(ia\phi_{B})\right\rangle $
which reads
\begin{equation}
{\mathcal{G}}_{B}(a)=\left(  \frac{\pi\mu}{\gamma(\beta^{2})}\right)
^{\dfrac{a^{2}}{2(1-\beta^{2})}}g_{0}(a)g_{S}(a)g_{A}(a)
\end{equation}
where
\begin{align}
\log g_{0}(a) &  =\int_{0}^{\infty}\frac{dt}{t}\left[  \frac{2\sinh^{2}(a\beta
t)\left(  e^{(1-\beta^{2})t/2}\cosh(t/2)\cosh(\beta^{2}t/2)-1\right)  }{\sinh
t\sinh(\beta^{2}t)\sinh((1-\beta^{2})t)}-a^{2}e^{-t}\right]  \nonumber\\
\log g_{S}(a) &  =\int_{0}^{\infty}\frac{dt}{t}\frac{\sinh^{2}(a\beta
t)\left(  2-\cos(2zt)-\cos(2z^{\ast}t)\right)  }{\sinh t\sinh(\beta^{2}%
t)\sinh((1-\beta^{2})t)}\\
\log g_{A}(a) &  =\int_{0}^{\infty}\frac{dt}{t}\frac{\sinh(2a\beta t)\left(
\cos(2zt)-\cos(2z^{\ast}t)\right)  }{\sinh t\sinh(\beta^{2}t)\cosh
((1-\beta^{2})t)}\nonumber
\end{align}
where the complex number $z$ is related to the parameters of the model
(\ref{bsing}) as
\begin{equation}
\cosh^{2}\pi z=\frac{\mu_{B}^{2}e^{-2i\beta\phi_{0}}}{\mu}\sin\pi\beta^{2}%
\end{equation}
and $z^{\ast}$ is the complex conjugate to $z$. The details and some
applications will be published elsewhere.

\item  In the next publication \cite{next} we will present an explicit
expression for the bulk-boundary structure constant (\ref{bbound}). Equation
(\ref{inversion}) then permits to resolve the system (\ref{bulk4}),
(\ref{bound4}) and (\ref{crossmatrix}) for the cross-matrix and obtain an
explicit expression for a special case of symmeric cross-matrix $K\left(
\begin{array}
[c]{cc}%
\alpha_{1} & \alpha_{1}\\
\alpha_{2} & \alpha_{2}%
\end{array}
\mid P,P^{\prime}\right)  $.

\item  The random lattice models of 2D quantum gravity allow in many cases to
find explicitely the partition functions of minimal models on fluctuating disk
with some bulk and boundary operators inserted \cite{Kostov1,Kostov2}.
Detailed comparison with the Liouville field thery predictions seems quite
interesting. The work on that is in progress.

\end{itemize}

\vspace{0.4cm}

\textbf{Aknowledgements}

The present study started as a common project with J.Teschner. In the
course of the project it turned out that the methods and 
results were rather complimentary, so that it was
decided to present the respective points of view in seperate
publications, see ref.\cite{PonsotTeschner1}. 

The work of V.F. and Al.Z. was partially supported by EU under contract
ERBFMRX CT 960012.

The work of A.Z. is supported in part by DOE grant \#DE-FG05-90ER40559.

\end{document}